\documentclass[preprint,10pt]{aastex}
 
\def\gs{\mathrel{\raise0.27ex\hbox{$>$}\kern-0.70em 
\lower0.71ex\hbox{{$\scriptstyle \sim$}}}}
\def\ls{\mathrel{\raise0.27ex\hbox{$<$}\kern-0.70em 
\lower0.71ex\hbox{{$\scriptstyle \sim$}}}}

\begin{document} 
\title{Photon vs Energy Magnitude Systems and the Measurement of
the Cosmological Parameters}

\author{Alex G. Kim \& Peter E. Nugent}
\affil{Lawrence Berkeley National Laboratory}
\affil{1 Cyclotron Rd., Berkeley, CA 94720}
\email{agkim@lbl.gov \& penugent@lbl.gov}

\begin{abstract}
The relative brightnesses of standard candles have long been known 
to be potentially powerful probes of distance.
The distance modulus, the difference between observed and absolute
magnitudes, has been associated with the values of the cosmological
parameters: Hubble's constant $H_0$,
the mass density $\Omega_M$ and the cosmological constant
$\Omega_\Lambda$.  In the literature the relationship between these
parameters and the distance modulus is calculated for
an energy magnitude system;
the Johnson-Cousins magnitude system used in observations is
in fact a photon-counting system.
In this paper, we present
the relation between observed and absolute photon magnitudes
in terms of the familiar energy distance modulus and derive
the correct form of the K-correction.
The differences between energy and photon systems are small relative
to the measurement errors of contemporary high-redshift supernova searches.
The distinction must be made, however,
for precision cosmological measurements such as those
planned for Type Ia supernovae.
\end{abstract}

\keywords{Distance scale --- supernovae: general}

\section{Introduction}
Measurements of the cosmological parameters using distance indicators
rely on the redshift-dependent evolution of the distance modulus $\mu$.
The distance modulus is measured as the difference between observed
and absolute magnitudes of a ``standard candle'' after K-correction
\citep{ok:kcorr} for
the redshifting of its spectrum.
The theoretical value for $\mu$ is related to the luminosity distance
$d_L(z)$ defined such that
a source with luminosity $L$ at redshift $z$ has observed 
energy flux $f$ as if the energy has been diluted to the surface
of a sphere with radius $d_L$, i.e.
$L=4 \pi d_L^2 f$ (e.g. \citet{carrollpressturner}).
Cosmological parameters can then be measured from their functional
dependence on $d_L$;
this technique has been used by two groups with
Type Ia supernovae [the High-$z$ team
\citep{riess_acc_98} and the Supernova Cosmology Project (SCP)
\citep{42SNe_98}] and gives evidence for an accelerating
universe.

Observations are in fact made with photon counters (CCD's, photo-multipliers)
and the luminosity distance is not the same as the
``photon luminosity
distance'' $d_\gamma$; if N is the photon luminosity and n is the
observed photon flux, then $N = 4 \pi d_\gamma^2 n$ where
$d_L=d_\gamma (1+z)^{1/2}$.  This has lead to some
confusion as to whether a ``photon'' distance
modulus should be used
to measure cosmological parameters, whether the magnitude system is
photon-based or energy-based, and which K-corrections should
be applied.
Such distinctions which previously have been unimportant are
significant as we move into an era of precision cosmology.
In this paper we rederive and expand upon the K-correction results of
\citet{1983ApJ...264..337S}.
We comment on the 
magnitude system and the Johnson-Cousins system in particular
(\S~\ref{mag:sec}).  We find that any ambiguity can be removed with
the proper definition of the K-correction for which we derive the
equations for both
photon and energy systems (\S~\ref{kcorr:sec}).  We conclude that
although the differences between the two K-corrections are small,
the distinction between energy and photon systems is important
for planned future high-precision supernova experiments (\S~\ref{con:sec}).

\section{Magnitude Systems}
\label{mag:sec}
The primary standards of a photometric system can have their magnitudes
measured either by their energy or photon flux ratios.
Unless a photon--energy conversion
correction is later applied, the flux system is determined by the detectors
used to measure the primaries.
The type of detector used in subsequent observations does not determine
whether the magnitude system is photon or energy based;
in principle the color and airmass corrections put observed magnitudes
into the primary system.

The Johnson-Cousins magnitude system prevalent today
is a photon system, what \citet{Johnson:1953} describe as
``a system of photoelectric photometry''.  As described in
\citet{Johnson:1951}, their observational setup employed
a photomultiplier as a detector, with the counts
being the number of ``deflections'' recorded by a potentiometer.
After an airmass correction these counts were directly converted to
magnitudes.
The secondary stars of
\citet{la:1973,la:1983,la:1992}
(whose raw data also were obtained with photon
counters) are calibrated via Johnson and Cousins primary standards and thus
must be in the photon system.  Observed magnitudes are therefore
photon-based and should be analyzed as such.

An illustrative example of where there is a numerical
difference between the two magnitude systems is a star that has the same
integrated $B$-band energy flux as Vega (which for simplicity
we consider to be the zero point of the magnitude system)
but has a different photon flux since it has a different spectral
energy distribution (SED).
Relative magnitude measurements with a single filter
of a set of stars with similar spectral energy distributions are independent
of whether we are photon counting or measuring energy; two stars
with the same SED but differing brightness will have
$$\Delta m = m_2^\gamma - m_1^\gamma = m_2^\epsilon - m_1^\epsilon \nonumber$$
where $m_1$ and $m_2$ are the stars' magnitudes.
It follows that
since the zeropoint of magnitude system is based on Vega,
the energy and photon magnitudes of A0V stars are identical: $m_{A0V}^\gamma =
m_{A0V}^\epsilon$.

As an aside, one
of the \citet{Johnson:1953} criteria for a photometric system
is
``a determination of the zero point of the color indices in terms of
a certain kind of star which can be accurately defined spectroscopically.''
Such knowledge, along with the shapes of the pass-band transmission functions,
do allow for calculated transformation
between photon and energy magnitude systems.  Indeed, much effort
has been placed in measuring and modeling the intrinsic SED of Vega
(\citet{dr:1980} and references therein).

\section{The K-correction}
\label{kcorr:sec}
We explicitly review the K-correction calculation of
\citet{kim_kcorr96} that has been used in SCP cosmological analysis. 
to remove any ambiguity.
We define the K-correction $K_{xy}$ such that
\begin{equation}
m_y^\alpha=M_x^\alpha + \mu(z) +K_{xy}^\alpha
\label{definition}
\end{equation}
where $\alpha=\{\gamma,\epsilon\}$ for photon or energy magnitude systems.
The observed magnitude in passband $y$ is $m_y$ and the absolute
magnitude in passband $x$ is $M_x$.
We adopt the theoretical expression
for the distance modulus, $\mu$,  based on luminosity distance.
In other words, the functional
form of $\mu(z;H_0,\Omega_M,\Omega_\Lambda)$ in
Equation~\ref{definition} is identical for
photon and energy systems.
Given $f_\lambda(\lambda)$ as the energy flux density of a supernova 10
parsecs away, we can compute the corresponding energy and photon
fluxes at high redshift.
\[
\begin{array}{lr}
f_\lambda(\lambda)d\lambda  & \mbox{Energy flux density in d$\lambda$ bin of a supernova 10
parsecs away}\\
n_\lambda(\lambda)d\lambda = \frac{\lambda d\lambda}{hc}f_\lambda(\lambda) & \mbox{Photon flux in d$\lambda$ bin of a supernova 10 parsecs away}\\
f^z_\lambda(\lambda)d\lambda  =  \frac{d\lambda}{1+z}f_\lambda\left(\frac{\lambda}{1+z}\right)\left(\frac{10 pc}{d_L(z)}\right)^2 & \mbox{Energy flux density in d$\lambda$ bin of a supernova at $z$}\\
n^z_\lambda(\lambda)d\lambda  =  \frac{\lambda d\lambda}{hc(1+z)}f_\lambda\left(\frac{\lambda}
{1+z}\right)\left(\frac{10 pc}{d_L(z)}\right)^2 & \mbox{Photon flux density in d$\lambda$ bin of a supernova at $z$}
\end{array}
\]
The $(1+z)^{-1}$ terms in the redshifted flux densities are due to
wavelength dilution \citep{ok:kcorr}.
The ratio between high and low-redshift photon flux is a factor
$1+z$ greater than the corresponding ratio for energy flux which suffers
from redshifted energy loss.  More precisely
\begin{equation}
\frac{n^z_\lambda(\lambda)}{n_\lambda(\lambda/(1+z))}=\frac{(1+z)f^z_\lambda(\lambda)}{f_\lambda(\lambda/(1+z))}.
\label{ratio}
\end{equation}
The fact that the relative photon fluxes of high-redshift supernovae are
$1+z$ ``brighter'' than energy fluxes can be interpreted as being due
to the latter's extra energy loss due to redshift. 

Using the fact that $\mu=-5\log{\left(\frac{10 pc}{d_L(z)}\right)}$ we can compute and
compare energy and photon K-corrections,
\begin{eqnarray}
  K_{xy}^\epsilon & = &  -2.5 \log
    \left(
    \frac
       {\int {\cal Z}^\epsilon_x(\lambda)S_x(\lambda)d\lambda}
       {\int {\cal Z}^\epsilon_y(\lambda)S_y(\lambda)d\lambda}
    \right)
    +2.5 \log(1+z) 
    +2.5 \log
    \left( 
    \frac
	{\int f_\lambda(\lambda)S_x(\lambda)d\lambda}
	{\int f_\lambda(\lambda/(1+z))S_y(\lambda)d\lambda}
    \right)
\label{ekcorr}
\end{eqnarray}
\begin{eqnarray}
  K_{xy}^{\gamma} & = & -2.5 \log
    \left(
    \frac
       {\int \lambda{\cal Z}^\gamma_x(\lambda)S_x(\lambda)d\lambda}
       {\int \lambda{\cal Z}^\gamma_y(\lambda)S_y(\lambda)d\lambda}
    \right)
    +2.5 \log(1+z) 
    +2.5 \log
    \left( 
    \frac
	{\int \lambda f_\lambda(\lambda)S_x(\lambda)d\lambda}
	{\int \lambda f_\lambda(\lambda/(1+z))S_y(\lambda)d\lambda}
    \right).
\label{gkcorr}
\end{eqnarray}
The filter transmission functions are given as $S_i(\lambda)$ where
$S_x$ is the rest-frame filter and $S_y$ is the observer filter.  (The
transmission functions give the fraction of photons transmitted at a
given wavelength where we assume no down-scattering.)  For the standard
star (i.e. calibrator) SED ${\cal Z}(\lambda)$ we assume the existence
of a standard star with identical properties as the supernova,
i.e. with exactly the same color and observed through the same
airmass.  Pragmatically, this assumption affirms perfect photometric
calibration to all orders of color and airmass.  For convenience, we
choose these secondary standards to have 0 magnitude.  In principle, a
different standard will be needed for each filter, choice of photon or
energy flux, and each source SED.  Each standard is labeled ${\cal
Z}_X^\alpha$ where $X=\{U,B,V,R,I,\ldots\}$ and
$\alpha=\{\gamma,\epsilon\}$ for photon or energy flux as defined
earlier.

Equations~\ref{ekcorr} and \ref{gkcorr} generalize the K-corrections
of \citet{1983ApJ...264..337S}\footnote{Note that in the
notation of \citet{1983ApJ...264..337S}, $f_\nu$ and $f_{\nu(1+z)}$
are the same function evaluated at different frequencies.}
and are precisely those given and
calculated in \citet{kim_kcorr96}.  In that paper, it was found that
the differences between the two K-corrections are non-zero but small,
$|K_{xy}^\epsilon-K_{xy}^{\gamma}|<0.07$ magnitudes.  They are a
function of redshift, filters, and supernova epoch and thus can cause
small systematic shifts in light-curve shapes and magnitude deviations
in the Hubble diagram.  The use of the incorrect K-correction will
have a significant effect on experiments with small $\ls 0.1$ targeted
magnitude errors.

\begin{figure}[h]
\plotone{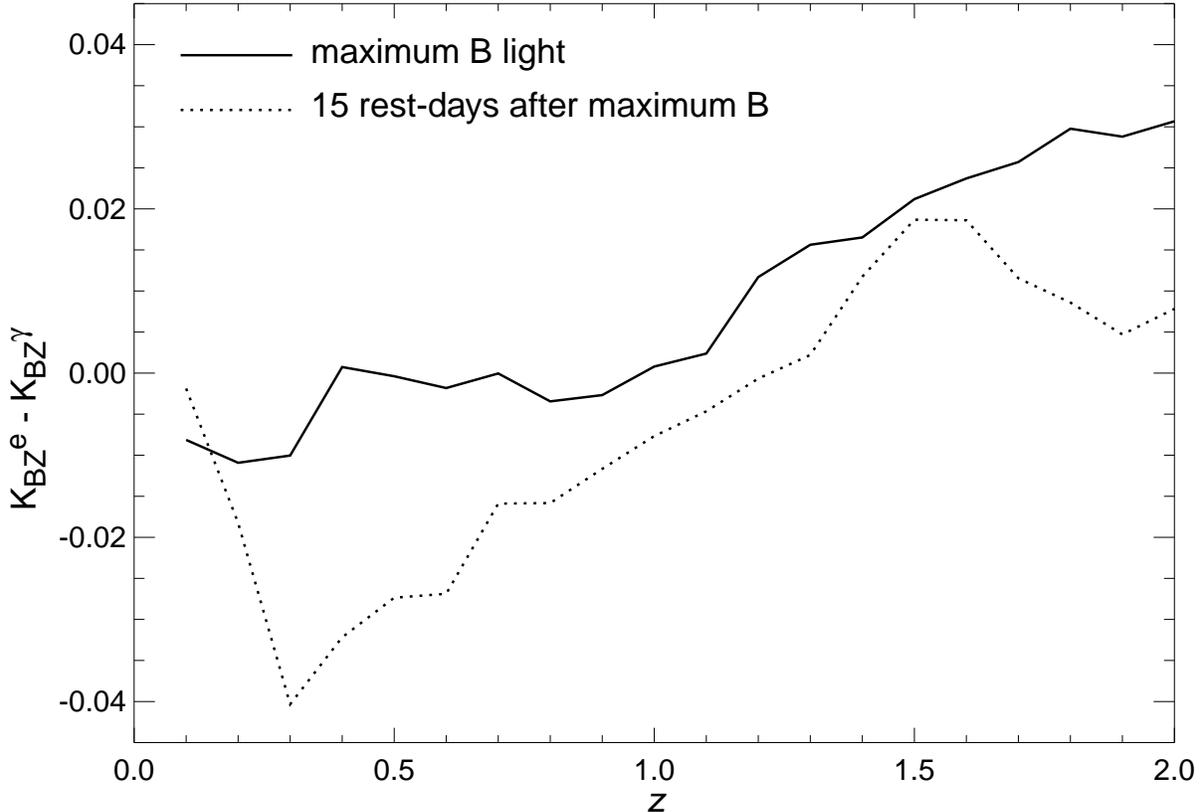}
\figcaption[fg1.eps]{$K_{BZ}^\epsilon - K_{BZ}^\gamma$ for a standard Type Ia
supernova at $B$ maximum and 15 days after maximum as a function of
redshift.  Measurements in $I$ and bluer filters for $z<1$ supernovae
and $j$ and redder filters for $z>1.5$ would provide a better match of
observed spectral regions.\label{fig}}
\end{figure}

To illustrate, in Figure~\ref{fig} we plot $K_{BZ}^\epsilon -
K_{BZ}^\gamma$ (where $Z$ refers to the passband and not redshift) for
a standard Type Ia supernova at $B$ maximum and 15 rest-frame days
after maximum out to $z=2$.  The differences are close to zero at $z
\sim 1.1$ where $B(\lambda/(1+z)) \sim Z(\lambda)$.  Beyond this
optimal redshift, the differences can be $>0.01$ magnitudes.  The
redder color of the supernova at the later epoch gives relatively larger
photon K-corrections over almost all redshifts.

The similarity in the two K-corrections is due to two competing terms
that nearly cancel.  A photon K-correction is $1+z$ brighter because
the supernova does not suffer redshifting energy loss.  However, the
zeropoint of the redder filter used to observe the redshifted
supernova is larger since an A0V photon spectrum is flatter than its
energy spectrum.  This makes the observed supernova magnitude
numerically fainter.  Consider the special case where
$S_y(\lambda)=S_x(\lambda/(1+z))$.  With perfect filter-matching the
specifics of the supernova spectrum are unimportant and the
K-corrections depend on the zeropoints:
\begin{eqnarray}
  K_{xy}^\epsilon & = &  -2.5 \log
    \left(
    \frac
       {\int {\cal Z}^\epsilon_x(\lambda)S_x(\lambda)d\lambda}
       {\int {\cal Z}^\epsilon_y(\lambda)S_y(\lambda)d\lambda}
    \right)\\
  K_{xy}^{\gamma} & = & -2.5 \log
    \left(
    \frac
       {(1+z)\int \lambda{\cal Z}^\gamma_x(\lambda)S_x(\lambda)d\lambda}
       {\int \lambda{\cal Z}^\gamma_y(\lambda)S_y(\lambda)d\lambda}
    \right)\\
 &=& -2.5 \log
    \left(
    \frac
       {(1+z) <\lambda_x>\int {\cal Z}^\gamma_x(\lambda)S_x(\lambda)d\lambda}
       {<\lambda_y>\int {\cal Z}^\gamma_y(\lambda)S_y(\lambda)d\lambda}
	\right)
\end{eqnarray}
where  $<\lambda>$ is the effective wavelength of the standard through
the filter.  As
long as the standard star is well behaved, we expect
the effective wavelength of the redshifted filter to be $1+z$ greater
then that of the restframe filter
$<\lambda_y> \sim (1+z)<\lambda_x>$ so that
\begin{equation}
	K_{xy}^{\gamma} \sim -2.5 \log
	\left(
    \frac
       {\int {\cal Z}^\epsilon_x(\lambda)S_x(\lambda)d\lambda}
       {\int {\cal Z}^\epsilon_y(\lambda)S_y(\lambda)d\lambda}
    \right) = K_{xy}^{\epsilon}.
\end{equation}
Choosing filters that accept the same spectral
region at both low and high redshifts not only reduce errors but
also reduces the difference between energy and photon K-corrections.

The effect of using
the ``energy'' distance modulus in defining the K-correction
in
Equations~\ref{definition}, \ref{ekcorr}, and \ref{gkcorr}
are seen in the open-filter K-corrections.
When $S_x=S_y=1$, the energy K-correction is unnecessary
and indeed $K_{xy}^\epsilon=0$. 
For the photon K-correction we find
$K_{xy}^\gamma=-2.5\log{(1+z)}$,
the difference between ``energy'' and ``photon'' distance moduli.

A simple measure for the  difference between single-filter K-corrections
is the ratio in effective wavelength
of a redshifted and unredshifted source
through that filter.   For example, sources with power-law SED's have
identical photon and energy K-corrections.
For low-redshift objects the difference in effective
wavelength should
be very small (unless they have pathological spectra) and thus make
little difference in distance determinations.  For example, a Type
Ia supernova at maximum at $z=0.1$ observed through the $B$-band
would have a distance modulus error of 0.02 magnitudes if the wrong
K-correction were applied.

\section{Conclusion}
\label{con:sec}
We have shown that the measurements $m_Y(z)-M_X$ do depend on whether the
magnitude system is based on energy or photon flux.  Although the 
``photon luminosity distance''
is shorter than the standard luminosity distance, we can still use the
relation $m_Y(z)=M_X+\mu(z)+K_{XY}$ with the appropriate definitions
of the K-corrections; the ones of \citet{kim_kcorr96} are
appropriate.  With this definition, the standard equations linking the
energy distance modulus to cosmology are
applicable.
The Johnson-Cousins magnitude system is in fact photon-based.  Therefore,
the $K^\gamma_{XY}$ K-correction should and has been used
in the supernova cosmology analysis of the Supernova Cosmology Project.
Although application of the incorrect K-correction would contribute
negligibly to the error budget of the current supernova sample, the
distinction is important for precision
experiments that require 0.02 magnitude
accuracies, such as the Supernova Acceleration Probe.
With the choice
of well-matched filters, differences between energy and photon K-corrections
can be minimal.

Using the ``count'' distance modulus
based on $d_\gamma$ in Equation~\ref{definition}
would provide a more physically satisfying definition of the count
K-correction.  Recall that $\mu^\epsilon= \mu^\gamma + 2.5\log{(1+z)}$. 
Then the extra $2.5\log{(1+z)}$ in the K-correction would give
\begin{equation}
  K_{xy}^{\gamma} = -2.5 \log
    \left(
    \frac
       {\int \lambda{\cal Z}^\gamma_x(\lambda)S_x(\lambda)d\lambda}
       {\int \lambda{\cal Z}^\gamma_y(\lambda)S_y(\lambda)d\lambda}
    \right) 
    +2.5 \log
    \left( 
    \frac
	{\int \lambda f_\lambda(\lambda)S_x(\lambda)d\lambda}
	{\int \lambda' f_\lambda(\lambda')S_y((1+z)\lambda')d\lambda'}
    \right).
\label{gkcorr2}
\end{equation}
In other words, the K-correction would depend simply on the
ratio of supernova photons in the rest-frame filter and a blue-shifted observer
filter, and the zeropoint.
This methodology would preserve the physical meanings that we
associate with both distance modulus and K-correction.
For simplicity, however, we here adopt the energy distance modulus
for both K-corrections to be consistent with the literature
and to ensure unambiguity when referring to
K-corrected magnitudes and distance moduli.

We would like to thank Arlo Landolt and Saul Perlmutter
for reading our manuscript and for their
insightful questions and comments.
This work was supported by the U.S. Department of Energy under contract No.
DE-AC03-76SF00098 and a NASA LTSA grant to PEN.


\end{document}